\documentstyle[preprint, revtex]{aps}
\def\beq{\begin{eqnarray}}
\def\eed{\end{eqnarray}}
\begin{document}
\draft

\begin{title}
\begin{center}
{Estimation of the Doping Dependence of Antiferromagnetism in the Copper
Oxide Material}
\end{center}
\end{title}
\author{Shiping Feng$^{1,2}$, Yun Song$^{1}$, and Zhongbing Huang$^{1}$}

\begin{instit}
$^{1*}$Department of Physics, Beijing Normal University, Beijing
 100875, China and \\
$^{2}$National Laboratory of Superconductivity, Academia Sinica,
 Beijing 100080, China
\end{instit}

\begin{abstract}
Within the $t$-$J$ model, we study the doping dependence of
antiferromagnetism in the copper oxide materials by considering quantum
fluctuations of spinons in the random-phase-approximation. The staggered
magnetization vanishes around doping $\delta=5\%$ for a reasonable
parameter value $t/J=5$, which is in agreement with the experiments
on copper oxide materials.
\end{abstract}
\pacs{75.10.Jm, 75.10.+x, 71.45.-d}

Since the discovery that the copper oxide sheets in the High Tc
superconductors show strong antiferromagnetic (AF) spin correlations
\cite{n1}, there has been increased interest in studying magnetic
properties of these systems. This followed
from the argument, made by Anderson and many other researchers \cite{n2},
that the essential physics of the copper oxide superconductors is
contained in doped antiferromagnets, where a central issue is the
relationship between the hole doping and AF spin correlations \cite{n3,n4}.
In particular, the phase diagrams of these materials as  functions  of
hole doping have been established by neutron scattering, muon spin
rotation, and magnetic resonance measurements \cite{n5}. When the hole
doping concentration exceeds some critical value (about 5$\%$),  AF
long-range order (AFLRO) disappears and the materials are converted
into nonmagnetic metals. It is believed \cite{n2} that the physics of
these materials may be effectively described by a 2D, large U Hubbard
model or its equivalent, the $t$-$J$ model.

In an attempt to understand the relationship between the hole doping and
AF spin correlations, many authors have studied holes moving in the
background of the spin resonating valence bond state \cite{n2}, spin
flux phase \cite{n6}, and spiral spin phase \cite{n7}. A hole in the AF
background has a very large effective mass because of frustrations
\cite{n10}. Numerical simulations suggest  \cite{n11} that the 2D
$t$-$J$ model is  N\' eel ordered at half-filling at temperature T=0,
and the suppression of AFLRO as a consequence of hole doping has
already been studied by using variational Monte-Carlo technique
\cite{n12}. Recently, the fermion-spin theory has been employed to
study the ground-state properties of the 2D $t$-$J$ model \cite{n13,n14}
within the mean-field-approximation (MFA). The results show that the
magnetized $\pi$-flux state, which is the coexistence of AFLRO with
a $\pi$-flux state, has the lowest energy at half-filling, and the AFLRO
is destroyed by hole doping of the order $\sim 10\% \sim 15\%$ for
reasonable values of the parameters. However, quantum fluctuations
are dropped completely in this mean-field treatment. In fact, there
is a strong coupling between spin (spinon) and charge (holon) degrees
of freedom \cite{n15}. It has been emphasized \cite{n2} that for small
spin values, strong quantum fluctuations of spinons may generate a novel
spin liquid ground-state. The purpose of this paper is to study these
issues within the $t$-$J$ model by considering quantum fluctuations of
spinons.

We start from the $t$-$J$ model on a square lattice which can be written
as
\begin{eqnarray}
H = -t\sum_{i\sigma,\eta >0}C^{\dagger}_{i\sigma}C_{i+\eta\sigma} + h.c.
- \mu \sum_{i\sigma}C^{\dagger}_{i\sigma}C_{i\sigma} +
J\sum_{i,\eta >0}{\bf S}_{i}\cdot {\bf S}_{i+\eta} ,
\end{eqnarray}
with $\eta =\hat{x}, \hat{y}$, $C^{\dagger}_{i\sigma}$ ($C_{i\sigma}$)
are the electron creation (annihilation) operators,
${\bf S}_{i}=C^{\dagger}_{i}{\bf \sigma} C_{i}/ 2$ are spin operators
with ${\bf \sigma}=(\sigma_{x},\sigma_{y},\sigma_{z})$ as Pauli
matrices, and $\mu$ is the chemical potential. The $t$-$J$ Hamiltonian
(1) is supplemented by the on-site local constraint,
$\sum_{\sigma}C^{\dagger}_{i\sigma}C_{i\sigma}\leq 1$, {\it i.e.},
there should be no doubly occupied sites.

The crucial requirement for the $t$-$J$ model is to impose this local
constraint for a proper understanding of the physical properties of
the copper oxide materials \cite{n16}. The local nature of the constraint
is of prime importance, and its violation may lead to unphysical results
\cite{n16}. To avoid these difficulties, a fermion-spin transformation
based on the charge-spin separation
\begin{eqnarray}
C_{i\uparrow}=h^{\dagger}_{i}S^{-}_{i},~~~~
C_{i\downarrow}=h^{\dagger}_{i}S^{+}_{i},
\end{eqnarray}
was proposed \cite{n13} to study the $t$-$J$ model,
where the spinless fermion operator $h_{i}$ keeps track of the charge
(holon) while the pseudospin operator $S_{i}$ keeps track of the spin
(spinon). In this case, the on-site local constraint for single occupancy
is satisfied even in the MFA. For discussing quantum spin problems within
the framework of fermion many particle theory, we map the quantum spin
onto the spinless fermion representation via the 2D Jordan-Wigner
\cite{n17} transformation, $S^{+}_{i}=f^{\dagger}_{i}e^{i\theta_{i}}$,
$S^{-}_{i}=f_{i}e^{-i\theta_{i}}$, with $f_{i}$ as a spinless fermion,
phase factor $\theta_{i}=\sum_{l\neq i}f^{\dagger}_{l}f_{l}B_{il}$, and
$B_{il}={\rm Imln}(Z_{l}-Z_{i})$, in the  complex representation of the
lattice sites $Z_{l}=X_{l}+iY_{l}$. In this case, it has been shown that
the $t$-$J$ model can be expressed as \cite{n13,n14,n18}
\begin{mathletters}
\begin{eqnarray}
H=H_{0}+H^{(1)}_{I} +H^{(2)}_{I},~~~~~~~~~~~~~~~~~~~~~~~~~~~~ \\
H_{0}=-\mu \sum_{i}h^{\dagger}_{i}h_{i}+{J_{eff}\over 2}
\sum_{i,\eta >0}(e^{i\int^{i+\eta}_{i}{\bf A}\cdot d{\bf l}}
f^{\dagger}_{i}f_{i+\eta}+e^{-i\int^{i+\eta}_{i}{\bf A}\cdot d{\bf l}}
f^{\dagger}_{i+\eta}f_{i}) \nonumber \\
-2J_{eff}\sum_{i}f^{\dagger}_{i}f_{i}
+{1\over 2}NJ_{eff}, ~~~~~~~~~~~~~~~~~~~~~~~~~~~~~~~~~\\
H^{(1)}_{I}=-t\sum_{i,\eta>0}h_{i}h^{\dagger}_{i+\eta}
(e^{i\int^{i+\eta}_{i}{\bf A}\cdot d{\bf l}}f^{\dagger}_{i}f_{i+\eta}+
e^{-i\int^{i+\eta}_{i}{\bf A}\cdot d{\bf l}}f^{\dagger}_{i+\eta}f_{i})
+h.c \\
H^{(2)}_{I}={1\over 2}\sum_{i,\eta >0}(2J_{eff})f^{\dagger}_{i}f_{i}
f^{\dagger}_{i+\eta}f_{i+\eta}, ~~~~~~~~~~~~~~~~~~~~~~~~~~~~~~~~~~
\end{eqnarray}
\end{mathletters}
where $N$ is the number of sites, and
$J_{eff}\approx J\langle (h_{i}h^{\dagger}_{i})(h_{i+\eta}
h^{\dagger}_{i+\eta})\rangle$, $H^{(1)}_{I}$ and $H^{(2)}_{I}$ are the
spinon-holon and spinon-spinon interaction Hamiltonians, respectively.
${\bf A}(r_{i})$ is the vector potential created by the 2D Jordan-Wigner
phase factor, which relates the 2D quantum spin problem to the fermion's
flux phase problem \cite{n6}. The Hamiltonian (3) contains both phase and
amplitude fluctuations. The phase fluctuations can be studied by using
the anyon and eikonal expansion techniques \cite{n19}. However, the
essential ingredients to give a qualitatively correct description of the
global ground-state properties perhaps are included in amplitude
fluctuations \cite{n18}. Therefore, in this paper, we only discuss the
amplitude fluctuations and  treat the phase factor in eq. (3) in MFA,
as in the previous works \cite{n13,n14,n18}. For convenience, we choose
the gauge field
\begin{eqnarray}
{\bf A}_{ij}={\pi\over 2}(y_{j}-y_{i})(-1)^{i}\hat{y},
\end{eqnarray}
 which is produced by a ${1\over 2}$ flux quantum
(or phase $\pi$) through each plaquette.

For discussing the self-consistent
spinon and holon Green's functions, we define the nearest-neighbor spin
bond-order amplitude, N\' eel order parameter, and holon particle-hole
parameter as
\begin{eqnarray}
\chi =\langle S^{+}_{i}S^{-}_{i+\eta}\rangle =
\langle e^{i\int^{i+\eta}_{i}{\bf A}\cdot d{\bf l}}f^{\dagger}_{i}
f_{i+\eta}\rangle , \\
\langle S^{Z}_{i}\rangle =(-1)^{i}M,~~~~~~~~~~~~~~~~\\
\phi =\langle h^{\dagger}_{i}h_{i+\eta}\rangle ,~~~~~~~~
\end{eqnarray}
respectively. This means we consider the coexistence of AFLRO with a
$\pi$-flux state, and this state, referred to as the magnetized
$\pi$-flux state, does not break the time-reversal symmetry and parity
\cite{n13,n14,n18}. The site subscripts of the order parameters $\chi$,
$\phi$, and $M$ have been dropped since the system is translation invariant.

Since there are two inequivalent sublattices A and B in the AF system,
the one-particle spinon and holon Matsubara Green's functions can be
defined as matrices,
\begin{mathletters}
\begin{eqnarray}
D({\bf k},\tau-\tau')=D_{L}({\bf k},\tau-\tau')
+ D_{T}({\bf k},\tau-\tau'), \\
D_{L}({\bf k},\tau-\tau')=-\left (\matrix{
\langle T_{\tau}f^{A}_{k}(\tau)f^{A\dagger}_{k}(\tau')\rangle &0\cr
0 &\langle T_{\tau}f^{B}_{k}(\tau)f^{B\dagger}_{k}(\tau')\rangle
\cr} \right ) , \\
D_{T}({\bf k},\tau-\tau')=-\left (\matrix{
0 &\langle T_{\tau}f^{A}_{k}(\tau)f^{B\dagger}_{k}(\tau')\rangle\cr
\langle T_{\tau}f^{B}_{k}(\tau)f^{A\dagger}_{k}(\tau')\rangle &0
\cr} \right ),
\end{eqnarray}
\end{mathletters}
and
\begin{mathletters}
\begin{eqnarray}
G({\bf k},\tau-\tau')=G_{L}({\bf k},\tau-\tau')
+ G_{T}({\bf k},\tau-\tau'), \\
G_{L}({\bf k},\tau-\tau')=-\left (\matrix{
\langle T_{\tau}h^{A}_{k}(\tau)h^{A\dagger}_{k}(\tau')\rangle &0\cr
0 &\langle T_{\tau}h^{B}_{k}(\tau)h^{B\dagger}_{k}(\tau')\rangle
\cr} \right ) , \\
G_{T}({\bf k},\tau-\tau')=-\left (\matrix{
0 &\langle T_{\tau}h^{A}_{k}(\tau)h^{B\dagger}_{k}(\tau')\rangle\cr
\langle T_{\tau}h^{B}_{k}(\tau)h^{A\dagger}_{k}(\tau')\rangle &0
\cr} \right ),
\end{eqnarray}
\end{mathletters}
where $\tau$ and $\tau'$ are the imaginary times, and
$\langle \ldots \rangle$ is an average over the ensemble.
$D_{L}(G_{L})$ and $D_{T}(G_{T})$ are the longitudinal and transverse
parts of the spinon (holon) Green's function
$D(G)$, respectively.

A self-consistent method of treating interaction Hamiltonians as
perturbation has been given in Ref. \cite{n20} within the Schwinger-boson
slave-fermion theory. We \cite{n18} have used similar self-consistent
calculations within the Jordan-Wigner approach to discuss the $t$-$J$ model
at half-filling and obtained some interesting results. In the following,
we will generalize the method in Ref. \cite{n18} to discuss the $t$-$J$
model away from half-filling within the fermion-spin approach.

With the gauge choice of eq. (4), the zero-order spinon and holon Green's
functions of $H_{0}$ in eq. (3b) are given by
\begin{eqnarray}
D^{(0)}_{L}(k)={i\omega_{n}+2J_{eff}
\over (i\omega_{n}+2J_{eff})^{2}-\omega^{2}_{0}({\bf k})}, ~~~~~
D^{(0)}_{T}(k)={J_{eff}(\sigma_{x}{\rm cos}k_{x}-\sigma_{y}{\rm cos}k_{y})
\over (i\omega_{n}+2J_{eff})^{2}-\omega^{2}_{0}({\bf k})},
\end{eqnarray}
and
\begin{eqnarray}
G^{(0)}_{L}(k)={1 \over i\omega_{n}+\mu}, ~~~~~~G^{(0)}_{T}(k)=0,
\end{eqnarray}
respectively, where $k=({\bf k},i\omega_{n})$, $\omega_{0}({\bf k})=
J_{eff}\Gamma_{{\bf k}}$, and
$\Gamma_{{\bf k}}=\sqrt{{\rm cos}^{2}k_{x}+{\rm cos}^{2}k_{y}}$.

According to the Dyson's theory \cite{n21}, the full spinon and holon
Green's functions of $H$ in eq. (3) due to considering the spinon-holon
and spinon-spinon interactions are formally expressed as,
\begin{eqnarray}
D(k)={1\over D^{(0)-1}(k)-\Sigma_{s}(k)}={1\over i\omega_{n}+2J_{eff}-
J_{eff}(\sigma_{x}{\rm cos}k_{x}-\sigma_{y}{\rm cos}k_{y})-
\Sigma_{s}(k)},~~~~
\end{eqnarray}
\begin{eqnarray}
G(k)={1\over G^{(0)-1}(k)-\Sigma_{h}(k)}={1\over i\omega_{n}+\mu
-\Sigma_{h}(k)}, ~~~~~~~~~~~~~~~
\end{eqnarray}
where $\Sigma_{s}(k)$ and $\Sigma_{h}(k)$ are the spinon and holon
self-energy functions, respectively. In the self-consistent calculation,
the spinon and holon Green's functions cannot be self-consistently
determined only by Eqs. (12) and (13), since the order parameters $\chi$,
$M$, and $\phi$ are also included in the spinon and holon self-energy
functions, and we must have another set of self-consistent equations to
determine these parameters. According to the definitions of the order
parameters in Eqs. (5)-(7) and the gauge choice of (4), we can
get the following equations,
\begin{eqnarray}
\chi={1\over 4}{2\over N}\sum_{{\bf k}(red)}{1\over \beta}
\sum_{i\omega_{n}}Tr[(\sigma_{x}{\rm cos}k_{x}-
\sigma_{y}{\rm cos}k_{y})D_{T}({\bf k},i\omega_{n})], \\
M={1\over 2}{2\over N}\sum_{{\bf k}(red)}{1\over \beta}
\sum_{i\omega_{n}}Tr[\sigma_{z}D_{L}({\bf k},i\omega_{n})], \\
\phi={1\over 2}{2\over N}\sum_{{\bf k}(red)}\gamma_{{\bf k}}
{1\over \beta}\sum_{i\omega_{n}}Tr[\sigma_{x}
G_{T}({\bf k},i\omega_{n})], \\
\delta={1\over 2}{2\over N}\sum_{{\bf k}(red)}{1\over \beta}
\sum_{i\omega_{n}}TrG_{L}({\bf k},i\omega_{n}),
\end{eqnarray}
where (red) means the summation is carried only over the reduced Brillouin
Zone. The order parameters $\chi$, $M$, $\phi$, and chemical potential
$\mu$ can be obtained by self-consistently solving Eqs. (14)-(17).

Although the matrix analysis is quite tedious, the calculation of the
first-order self-energy (Hartree-Fock self-energy) function is
straightforward. Following the discussions in Ref. \cite{n18}, we obtain
the spinon Hartree-Fock self-energy
\begin{eqnarray}
\Sigma^{(HF)}_{sL}(k)=2J_{eff}-m\rho\sigma_{z}, ~~~\Sigma^{(HF)}_{sT}(k)
=2(\phi t- \chi J_{eff})(\sigma_{x}{\rm cos}k_{x}-\sigma_{y}{\rm cos}k_{y}),
\end{eqnarray}
and the holon Hartree-Fock self-energy
\begin{eqnarray}
\Sigma^{(HF)}_{hL}(k)=0, ~~~~~~\Sigma^{(HF)}_{hT}(k)=
\epsilon_{{\bf k}}\sigma_{x},
\end{eqnarray}
where $\rho =(1-2\chi)J_{eff}+2t\phi$, $m={4MJ_{eff}\over \rho}$,
$\epsilon_{{\bf k}}=4\chi t\gamma_{{\bf k}}$, and
$\gamma_{{\bf k}}={1\over 2}({\rm cos}k_{x}+{\rm cos}k_{y})$.
$\Sigma^{(HF)}_{sL}(\Sigma^{(HF)}_{hL})$ and $\Sigma^{(HF)}_{sT}
(\Sigma^{(HF)}_{hT})$ are the longitudinal and transverse parts of the
spinon (holon) Hartree-Fock self-energy $\Sigma^{(HF)}_{s}
(\Sigma^{(HF)}_{h})$, respectively. Substituting
these results of $\Sigma^{(HF)}_{sL}(k)$, $\Sigma^{(HF)}_{sT}(k)$,
$\Sigma^{(HF)}_{hL}(k)$, and $\Sigma^{(HF)}_{hT}(k)$ into Eqs. (12)
and (13), we obtain self-consistent Hartree-Fock (SCHF) spinon and
holon Green's functions,
\begin{eqnarray}
D^{(HF)}_{L}(k)={i\omega_{n}-m\rho\sigma_{z}\over (i\omega_{n})^{2}-
\omega^{2}({\bf k})}, ~~~~~~ D^{(HF)}_{T}(k)={\rho (\sigma_{x}{\rm cos}
k_{x}-\sigma_{y}{\rm cos}k_{y})
\over (i\omega_{n})^{2}-\omega^{2}({\bf k})},
\end{eqnarray}
and
\begin{eqnarray}
G^{(HF)}_{L}(k)={i\omega_{n}\over (i\omega_{n})^{2}-\epsilon^{2}_{{\bf k}}},
~~~~~~G^{(HF)}_{T}(k)={\epsilon_{{\bf k}} \sigma_{x}\over (i\omega_{n})^{2}
-\epsilon^{2}_{{\bf k}}},
\end{eqnarray}
respectively, where $\omega({\bf k})=\rho E_{{\bf k}}$ and
$E_{{\bf k}}=\sqrt{\Gamma^{2}_{{\bf k}}+m^{2}}$. This SCHF solution is
exactly the usual self-consistent MFA in the previous works \cite{n13,n14}.
At half-filling, the staggered magnetization and ground-state energy
are $M=0.389$ and $E_{0}=-0.33J$ per bond, respevtively. Away from
half-filling, AFLRO vanishes around doping $\delta=10-15\%$ for reasonable
values of the parameter $t/J$, which are in rough agreement with
experiments \cite{n5} and Monte-Carlo simulations \cite{n12}. In this
case, the SCHF Hamiltonian describes free spinons and holons moving in
an average Chern-Simons gauge field of flux $\pi$ per plaquette \cite{n13}.

The mean-field phase boundary between AFLRO and disordered states is,
of course, at somewhat higher doping $\delta$ than the value given by
experiments and numerical simulations. The $t$-$J$ model is
characterized by a competition between the kinetic energy ($t$)
and magnetic energy ($J$). The magnetic energy $J$ favors an AFLRO for
spins, whereas the kinetic energy $t$ favors delocalization of holes
and tends to destroy AFLRO. According to the experiments \cite{n5}, the
suppression of AFLRO depends on both hole dopings and quantum fluctuations
of spinons. Therefore in this paper the holon part will be restricted to
the Hartree-Fock level, and we consider quantum fluctuations of spinons
due to the spinon-spinon interaction $H^{(2)}_{I}$ in Eq. (3d) within the
random-phase-approximation (RPA).

The analytic spinon Green's function is very complicated as emphasized
in Refs. \cite{n18,n20}. To make  the actual calculation feasible, we
start from the SCHF solution, {\it i.e.}, the full spinon Green's
function is replaced by the spinon Hartree-Fock Green's function in
the calculation of the spinon RPA self-energy \cite{n18,n20}. The
detailed process to obtain the spinon RPA Green's function within the
$t$-$J$ model at half-filling has been given in Ref. \cite{n18}.
Following the pocedure described there\cite{n18}, we get the spinon
RPA Green's function in the present case,
\begin{eqnarray}
D^{(RPA)}(k)={1\over D^{(HF)-1}(k)-\Sigma^{(RPA)}(k)} ~~~~~~~~~~~~~
~~~ \nonumber \\
={i\omega_{n}-[m\rho-A({\bf k})]\sigma_{z}+[\rho+B({\bf k})]
(\sigma_{x}{\rm cos}k_{x}-\sigma_{y}{\rm cos}k_{y})\over
(i\omega_{n})^{2}-\omega^{2}_{RPA}({\bf k})} ,
\end{eqnarray}
where $\omega_{RPA}({\bf k})=\sqrt{[m\rho-A({\bf k})]^{2}+
[\rho+B({\bf k})]^{2}\Gamma^{2}_{{\bf k}}}$, and
\begin{mathletters}
\begin{eqnarray}
A({\bf k})=2J^{2}_{eff}({2\over N})^{2}\sum_{{\bf pq}(red)}
{1+{\rm cos}(p_{x}-p_{y})\over \varepsilon_{RPA}({\bf p})}
(1-{m^{2}\over E_{{\bf q}}E_{{\bf p+q}}}) ~~~~~~~~~~~~\nonumber \\
\times {{m\over E_{{\bf k+p}}}
(\omega_{{\bf q}}+\omega_{{\bf p+q}}+
\omega_{{\bf k+p}})+\omega_{{\bf k}}
\over (\omega_{{\bf q}}+\omega_{{\bf p+q}}+\omega_{{\bf k+p}})^{2}
+\omega^{2}_{{\bf k}}},~~ \\
B({\bf k})=2J^{2}_{eff}({2\over N})^{2}\sum_{{\bf pq}(red)}
{1+{\rm cos}(p_{x}-p_{y})\over \varepsilon_{RPA}({\bf p})}
{\gamma_{{\bf p}}\over E_{{\bf k+p}}}{\Gamma^{2}_{{\bf q}}
\over E_{{\bf q}}E_{{\bf p+q}}} ~~~~~~~~\nonumber \\
\times {(\omega_{{\bf q}}+\omega_{{\bf p+q}}+
\omega_{{\bf k+p}})+\omega_{{\bf k}}\over (\omega_{{\bf q}}+
\omega_{{\bf p+q}}+\omega_{{\bf k+p}})^{2}+\omega^{2}_{{\bf k}}}.~~
\end{eqnarray}
\end{mathletters}
Here $\varepsilon_{RPA}(p)$ is the spinon RPA dielectric function.
With the above holon and spinon Green's functions, we obtain the
self-consistent equations at zero temperature ,
\begin{eqnarray}
\chi=-{1\over 4}{2\over N}\sum_{{\bf k}(red)}{[\rho+B({\bf k})]
\Gamma^{2}_{{\bf k}}\over \omega_{RPA}({\bf k})}, \\
M={2\over N}\sum_{{\bf k}(red)}{m\rho-A({\bf k})\over
2\omega_{RPA}({\bf k})}, \\
\phi={2\over N}\sum_{{\bf k}(red)}{\gamma_{{\bf k}}\over 2}
[\theta(\mu-\epsilon_{{\bf k}})-\theta(\mu+\epsilon_{{\bf k}})], \\
\delta={2\over N}\sum_{{\bf k}(red)}{1\over 2}
[\theta(\mu-\epsilon_{{\bf k}})+\theta(\mu+\epsilon_{{\bf k}})],
\end{eqnarray}
where the effect of quantum fluctuations of spinons from the
spinon-spinon interaction have been included in $A({\bf k})$,
$B({\bf k})$, and $\omega_{RPA}({\bf k})$, which will affect the
parameters $\chi$, $M$, and $\mu$, but  not the parameter $\phi$ at low
dopings. This is because  the effect due to the variance of parameter
$\chi$ in $\epsilon_{{\bf k}}=4\chi t\gamma_{{\bf k}}$ in Eqs. (26) and
(27) can be incorporated into the parameter $t/J$. On the other hand,
we have shown
in Ref. \cite{n13} that at low dopings the holon particle-hole order
parameter $\phi$ increases roughly linearly upon doping, and is almost
independent of $t/J$. Therefore in this case we can take the mean-field
values of the parameter $\phi$ to calculate the RPA self-consistent
equations (24) and (25) for the parameters $\chi$ and $M$. We have
performed such RPA calculation. At half-filling, the result is the
same as those in Ref. \cite{n18}, $i.e.$ the staggered magnetization
and RPA ground-state energy are $M=0.327$ and $E_{0}=-0.332J$ per bond,
respectively, which are very close to the best numerical result \cite{n23}
$M=0.31$ and $E_{0}=-0.3346J$. Away from half-filling, AFLRO is quickly
destroyed by doping of the order of $\delta=5\%$ for a reasonable
parameter value $t/J=5$,
which is plotted in Fig. 1 (solid triangles). For comparison,
the experimental measurements on $(La_{1-x}Ba_{x})_{2}CuO_{4}$ (solid
squares, the dashed line is guide for eyes) \cite{n5} and numerical
results (solid circles) \cite{n12} are also plotted in Fig. 1. From Fig. 1,
it is obvious that our result is in agreement with the experiments
\cite{n5} and Monte-Carlo simulations \cite{n12}. As compared with the
mean-field result \cite{n13}, the present RPA results indicate that both
hole dopings and quantum fluctuations of spinons lead to a strong
suppression of AFLRO.

In summary, we have studied the doping dependence of AFLRO in the copper
oxdie materials within the $t$-$J$ model. We have performed a
self-consistent calculation, where the holon part is restricted to the
Hartree-Fock level, and the quantum fluctuation of spinons is considered
within the RPA. We have considered only the amplitude fluctuations.
AFLRO vanishes around doping $\delta=5\%$ for a
reasonable parameter value $t/J=5$, which is in agreement with the
experiments on copper oxide materials \cite{n5} and Monte-Carlo
simulations \cite{n12}.

\acknowledgments
The authors would like to thank Prof. X. Xu and Prof. Z. X. Zhao for helpful
discussions. This work is supported by the National Science Foundation
Grant No. 19474007 and the Trans-Century Training Programme Foundation
for the Talents by the State Education Commission of China.

\references{

\bibitem [*] {add} Mailing address.

\bibitem {n1} D. Vaknin {\it et al.}, Phys. Rev. Lett. {\bf 58}, 2802
(1987).

\bibitem {n2} P. W. Anderson, Science {\bf 235}, 1196 (1987); F. C. Zhang
and T. M. Rice, Phys. Rev. B{\bf 37}, 3759 (1988).

\bibitem {n3} A. P. Kampf, Phys. Rep. {\bf 249}, 219 (1994), and
references therein.

\bibitem {n4} See, {\it e.g.}, "{\it High Temperature Superconductivity}",
Proc. Los Alamos Symp., 1989, K. S. Bedell {\it et al.}, eds.
(Addison-Wesley, Redwood City, California, 1990).

\bibitem {n5} Y. Kitaoka {\it et al.}, Physica C{\bf 153-155}, 733 (1988);
J. H. Brewer {\it et al.}, Phys. Rev. Lett. {\bf 60}, 1073 (1988); D. W.
Cooke {\it et al.}, Phys. Rev. B{\bf 52}, R3864 (1995).

\bibitem {n6} I. Affleck and J. B. Marston, Phys. Rev. B {\bf 37},
3774 (1988).

\bibitem {n7} B. Shraiman and E. Siggia, Phys. Rev. Lett.
{\bf 62}, 1564 (1989).

\bibitem {n10} C. L. Kane, P. A. Lee, and
N. Read, Phys. Rev. B {\bf 39}, 6880 (1989); Z. B. Su, Y. M. Li, W.
Y. Lai, and L. Yu, Phys. Rev. Lett. {\bf 63}, 1318 (1989).

\bibitem {n11} S. Liang, B. Doucot, and P. W. Anderson, Phys. Rev. Lett.
{\bf 61}, 365 (1988).

\bibitem {n12} T. K. Lee and Shiping Feng, Phys. Rev. B {\bf 38},
11809 (1988).

\bibitem {n13} Shiping Feng, Z. B. Su, and L. Yu, Phys. Rev. B
{\bf 49}, 2368 (1994); Mod. Phys. Lett. B{\bf 7}, 1013 (1993).

\bibitem {n14} Shiping Feng, Physica C{\bf 232}, 119 (1994); X. Xu,
Y. Song, and Shiping Feng, Mod. Phys. Lett. B{\bf 9}, 1623 (1995);
X. Xu, Y. Song, and Shiping Feng, Acta Physica Sinica, {\bf 45},
1390 (1996).

\bibitem {n15} See, {\it e. g.}, the review, L. Yu, in "{\it Recent
Progress in Many-Body Theories}", vol. 3, T. L. Ainsworth {\it et al.},
eds.( Plenum, 1992) P.157.

\bibitem {n16} Shiping Feng, J. B. Wu, Z. B. Su, and L. Yu,
Phys. Rev. B{\bf 47}, 15192 (1993); L. Zhang, J. K. Jain, and
V. J. Emery, Phys. Rev. B{\bf 47}, 3368 (1993).

\bibitem {n17} E. Mele, Phys. Scripta {\bf T27}, 82 (1988);
E. Fradkin, Phys. Rev. Lett. {\bf 63}, 322 (1989); G. Baskaran,
in "{\it Two-Dimensional Strongly Correlated Electronic Systems}",
Z. Z. Gan and Z. B. Su, eds.(Gordon and Breach), 1989, p.72;
Y. R. Wang, Phys. Rev. B {\bf 46}, 151 (1992).

\bibitem {n18} Shiping Feng, Phys. Rev. B{\bf 53}, 11671 (1996).

\bibitem {n19} A. M. Tikofsky and R. B. Laughlin, Phys. Rev. B
{\bf 50}, 10165 (1994); D. V. Khveshchenko and P. C. E. Stamp,
Phys. Rev. Lett. {\bf 71}, 2118 (1993).

\bibitem {n20} Y. M. Li {\it et al.}, Phys. Rev. B {\bf 45}, 5428
(1992).

\bibitem {n21} G. D. Mahan, {\it Many-Particle Physics } (Plenum, New
York, 1981).

\bibitem {n23} N. Trivedi and D. M. Ceperley, Phys. Rev. B {\bf 40},
2747 (1989).

}

\figure{Calculated values of the staggered magnetization of the $t$-$J$
model as a function of the hole concentration $\delta$ for $t/J$=5 (solid
triangles). The solid squares and solid circles correspond to the
experimental measurements on $(La_{1-x}Ba_{x})_{2}CuO_{4}$ \cite{n5} and
numerical results \cite{n12}, respectively. $M_{0}$ is the value of the
staggered magnetization at half-filling.}

\end{document}